\documentclass[12pt,preprint]{aastex}

\shorttitle{Analysis and Modeling of Two Flare Loops}

\shortauthors{Li, Qiu, \& Ding}

\begin{document}

\title{Analysis and Modeling of Two Flare Loops \\Observed by AIA and EIS}

\author{Y. Li$^{1,2,3}$, J. Qiu$^{2}$, M. D. Ding$^{1,3}$}
\affil{$^1$School of Astronomy and Space Science, Nanjing University, Nanjing 210093, China}
\affil{$^2$Department of Physics, Montana State University, Bozeman, MT 59717, USA}
\affil{$^3$Key Laboratory for Modern Astronomy and Astrophysics (Nanjing University), Ministry of Education, Nanjing 210093, China}
\email{}

\begin{abstract}

We analyze and model an M1.0 flare observed by \textit{SDO}/AIA and \textit{Hinode}/EIS to investigate how flare loops are heated and evolve subsequently. The flare is composed of two distinctive loop systems observed in EUV images. The UV 1600 \AA\ emission at the feet of these loops exhibits a rapid rise, followed by enhanced emission in different EUV channels observed by AIA and EIS. Such behavior is indicative of impulsive energy deposit and the subsequent response in overlying coronal loops that evolve through different temperatures. Using the method we recently developed, we infer empirical heating functions from the rapid rise of the UV light curves for the two loop systems, respectively, treating them as two big loops of cross-sectional area 5\arcsec\ by 5\arcsec, and compute the plasma evolution in the loops using the EBTEL model \citep{klim08}. We compute the synthetic EUV light curves, which, with the limitation of the model, reasonably agree with observed light curves obtained in multiple AIA channels and EIS lines: they show the same evolution trend and their magnitudes are comparable by within a factor of two. Furthermore, we also compare the computed mean enthalpy flow velocity with the Doppler shift measurements by EIS during the decay phase of the two loops. Our results suggest that the two different loops with different heating functions as inferred from their footpoint UV emission, combined with their different lengths as measured from imaging observations, give rise to different coronal plasma evolution patterns captured both in the model and observations.

\end{abstract}

\keywords{Sun: corona -- Sun: flare -- Sun: magnetic reconnection -- Sun: UV radiation}

\section{Introduction}

It is widely accepted that magnetic reconnection occurs in solar flares with a large amount of energy released in the corona \citep{prie02}. By reconnection, magnetic field lines change connectivity and form new loops. The released energy is transported downward along the loop by non-thermal electrons or thermal conduction to produce enhanced optical and ultraviolet (UV) emissions at the footpoints of the loop. In the meanwhile, impulsive energy deposition in the lower atmosphere evaporates hot plasma into the newly formed flare loops. The plasma then cools down, becoming visible sequentially as loops in soft X-ray and extreme-UV (EUV) emissions.

Assuming that energy deposit takes place during the impulsive phase of the flare, some modelers adopt a loop heating rate proportional to the observed hard X-ray light curve and model heating and evolution of the flare as one single loop (e.g., \citealt{fish90, raft09}). However, it has been recognized that most flares consist of multiple loops formed and heated successively \citep{hori98, asch05, mulu11}. With this consideration, \cite{warr06} applied multi-thread 1D hydrodynamic simulations to a flare, and was able to reproduce the soft X-ray radiation of the flare as observed by \textit{GOES} and \textit{Yohkoh}. To match observations, the author used 50 threads (flare loops), each heated at a different time for about 200 s by a prescribed heating rate. Similarly, \cite{hock12} have modeled a two-phase flare observed by the Atmospheric Imaging Assembly (AIA; \citealt{leme12}) and EUV Variability Experiment (EVE; \citealt{wood12}) on board the \textit{Solar Dynamics Observatory} (\textit{SDO}). They observed different flare loop systems brightened in these two phases, and used the 0D model called ``enthalpy-based thermal evolution of loops" (EBTEL; \citealt{klim08}) to compute plasma evolution in these loops with arbitrary heating rates that are adjusted to match with the observed EUV radiation. In their study, 22 loops are heated throughout the flare to match the observed flare evolution.

Key to all loop heating models is the heating function, which describes where, when, for how long, and by how much a flare loop is heated. To date, most multi-thread loop heating models prescribe an arbitrary number of loops, each given an arbitrary amount of heating at an arbitrary time, and the parameters are adjusted to best match the observed light curves of coronal emission. To relate flare heating with magnetic energy release in a more realistic manner, \cite{long10} developed a model to compute the rate of energy released in reconnection using the time-dependent reconnection flux measured from observations of flare ribbon evolution. The amount of released energy is then used to compute flare plasma evolution and coronal radiation and compare with observations by \textit{GOES} and \textit{RHESSI}. Recently, \cite{qiuj12} introduced an alternative method to observationally infer heating rates by analyzing spatially resolved UV light curves emitted at the feet of over a thousand flare loops anchored to brightened UV footpoints. The loop heating occurs when the UV emission starts to rise, the duration of heating is twice the rise time of the UV light curve, and the magnitude of the heating rate is assumed to be directly proportional to the UV counts measured at the foot of the loop. With this method, many properties of the heating rate are observationally constrained.

For its high efficiency in modeling a large number of loops, a number of studies have used the EBTEL model to compute plasma evolution in flare loops, as listed in Table \ref{comtab}. These studies by different groups treat the model inputs and outputs in different ways. The input to the model includes the number of flare loops and their properties (such as the loop length) and their heating functions. The model-calculated temperature and density of flare loops are then used to compute X-ray and EUV emissions by flare loops that can be compared with observations. Furthermore, the EBTEL model includes a mean enthalpy flux between the corona and its base at the transition region, which describes upflows predominant during the heating phase and downflows during the cooling phase. This is consistent with the picture of chromospheric evaporation in the early phase and then coronal condensation in the decay phase of the flare, as reported in observations \citep{anto83,anto85,anto90,schm87,ding96,bros03, raft09} and 1D hydrodynamic models \citep{dosc82, fish85a, fish85b}. Therefore, the EBTEL computed flow can be also compared with Doppler shift measurements to provide another constraint to flare heating models. 

\begin{table}[h]
\centering
\tiny
\caption{\textsc{\small{Comparison among the studies referring to the EBTEL model and flare heating}}}
\begin{tabular}{cccc}
\tableline
\tableline
\multicolumn{1}{c}{studies} & heating function    & loop  number    & compare with observations \\
                                                &                                  &                            & (instruments) \\
\tableline
\tableline
  \cite{raft09}    & inferred from hard X-ray light curve                & 1 & temperature, emission measure, velocity\\
                           &                                                                             &    & (\textit{RHESSI}, \textit{GOES}, \textit{TRACE}, CDS)  \\
\tableline                            
  \cite{hock12} & free parameters                                                 & 22 & EUV light curves\\
                           &                                                                             &      & (EVE)\\
\tableline
  \cite{qiuj12}    & inferred from spatially resolved UV light curves    & thousands & X-ray and EUV light curves\\  
                           &                                                                                   &                    & (\textit{GOES}, AIA) \\                          
\tableline
    this study       & inferred from spatially resolved UV light curves     & 2                  & EUV light curves, velocity\\
                           &                                                                                    &                     &(AIA, EIS)  \\    
\tableline
\end{tabular}  
\label{comtab}  
\end{table}

Following \cite{qiuj12}, we will observationally infer heating functions from UV light curves and compute plasma evolution in individual flare loops. It, however, takes a few steps ahead of previous studies by making use of the most stringent observational constraints, in both the input heating functions and the output radiation and flow signatures. The novel aspects in this study include (1) the heating functions and predicted coronal radiations are derived and compared for different loops that are identified from imaging EUV observations and exhibit distinctive evolution patterns; (2) both radiation and velocity signatures are compared with observations in multiple lines by multiple instruments, AIA and the EUV Imaging Spectrometer (EIS; \citealt{culh07}) on board \textit{Hinode}. Such more comprehensive analysis and modeling allow us to better constrain the loop heating model and diagnose flare heating more accurately. Whereas \cite{qiuj12} have used EBTEL to model a few thousand loops with the same set of free parameters, in this study, we will focus on modeling and analyzing only two loops in order to refine determination of heating functions. The 0D EBTEL allows us to investigate more efficiently variations of these free parameters and gain insight into the underlying physics that governs these parameters. Among these parameters, one is related to the heating rate, and the other describes the energy loss through the lower atmosphere. These parameters might not be the same in different loop systems, and will be examined in the present study. It also should be noted that the 0D EBTEL code is bench-marked by more sophisticated 1D hydrodynamical simulations with reasonable agreement \citep{klim08}; on the other hand, with the EBTEL model combined with our method, we will be able to determine the first-order heating rates to be used as a valuable reference for more sophisticated 1D hydrodynamic models. In the following Section \ref{obs}, we describe flare observations. The model computation and comparison with observations are presented in Section \ref{result}.  We give the summary and discussions in Section \ref{discon}.

\section{Observations and Data Reduction}
\label{obs}

\begin{figure*}
\begin{center}
\includegraphics[height=17cm]{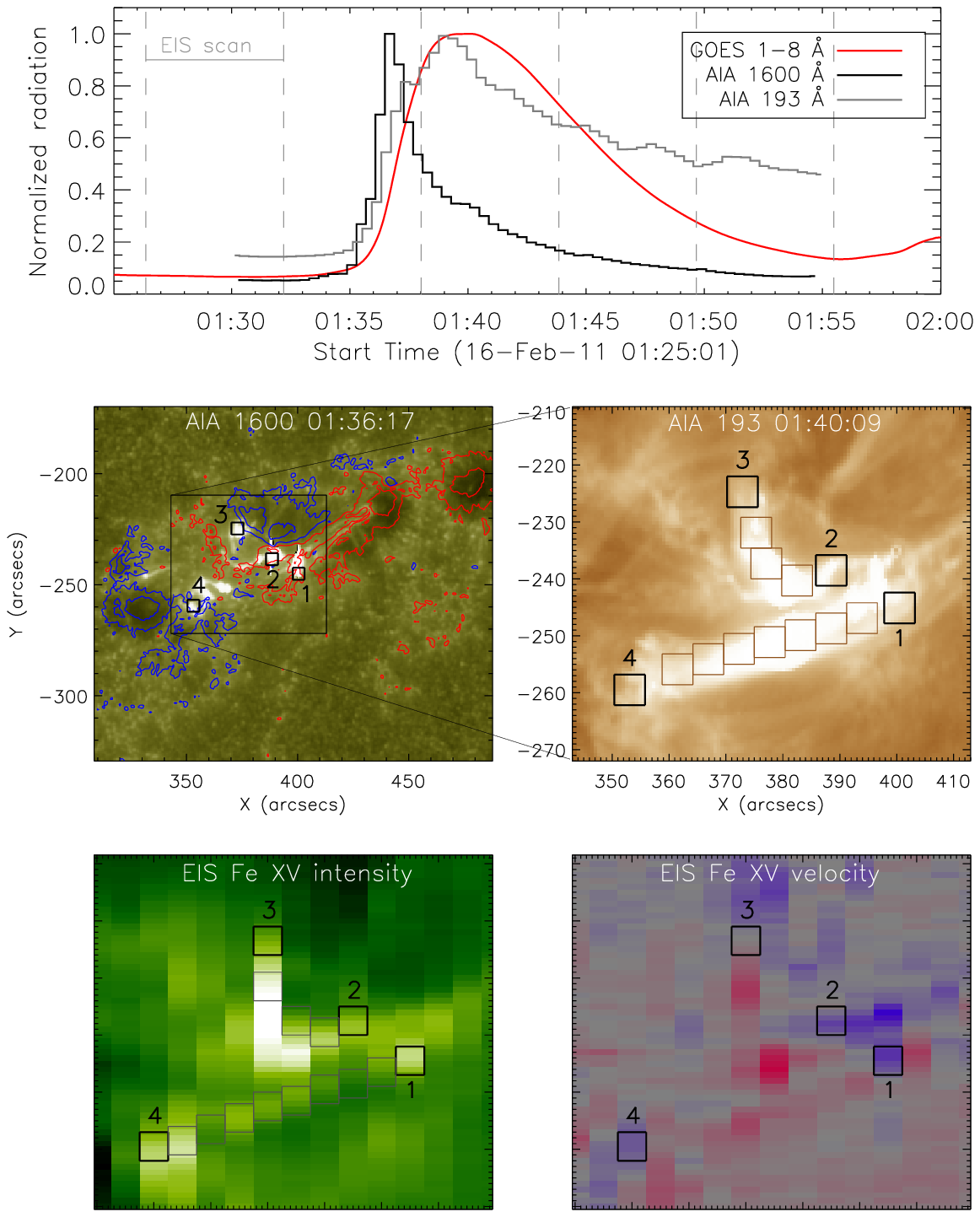}
\caption{Top panel: normalized light curves of the 2011 February 16 M1.0 flare in \textit{GOES} 1-8 \AA, AIA 1600 \AA, and AIA 193 \AA. The vertical dashed lines denote the time ranges of EIS five scans over the flaring region. Middle panels: AIA 1600 \AA~and 193 \AA~images of the flare. The red and blue contours in the left panel refer to the positive and negative magnetic polarities, respectively. The right panel zooms in the box region of the left panel. The four small black boxes mark four footpoints, and the series of small grey boxes are used to measure the total emission from the loop. Each box has a cross-sectional area of $5\arcsec\times5\arcsec$. Bottom panels: EIS Fe {\sc xv} intensity and Doppler velocity maps for the box region around the \textit{GOES} 1-8 \AA~peak time.}
\label{obsfig}
\end{center}
\end{figure*}

Observations of a \textit{GOES} M1.0 class flare on 2011 February 16 are summarized in Figure \ref{obsfig}. The top panel shows normalized light curves of the flare in \textit{GOES} 1-8 \AA, UV 1600 \AA, and EUV 193 \AA~bands. UV and EUV images in the middle panels were obtained by AIA with the pixel scale of 0.6\arcsec~and time cadences of 24~s and 12~s, respectively. Seen from the figure, the flare exhibits a rapid rise in the UV 1600 \AA. The UV emission is from the upper-chromosphere or transition region, where flare loops are anchored. It is followed by \textit{GOES} soft X-ray emission, and then the EUV emission at 193 \AA. This pattern indicates an impulsive heating of flare plasma and subsequent cooling toward lower temperatures. From AIA imaging observations, we can identify flare loops and their footpoints to study the heating and evolution of individual flare loops.

The flare was also observed by EIS. The 2\arcsec~slit of EIS scanned the flaring region from west to east using a 5\arcsec~step mode with an interval of 5 minutes and 50 seconds. The time ranges of five scans during the flare are indicated by the vertical dashed lines in the top panel of Figure \ref{obsfig}. These scans cover a field of view of $180\arcsec\times160\arcsec$ as indicated by the black box in the middle-left image in Figure \ref{obsfig}. The EIS data is reduced using the standard EIS software data reduction package. The software corrects detector bias and dark current, as well as hot pixels and cosmic ray hits, and yields absolute intensities in units of erg~cm$^{-2}$~s$^{-1}$~sr$^{-1}$~\AA$^{-1}$. In this study, we focus on analyzing the Fe {\sc xii} 195.12 line (formed at $\sim$1.25~MK) and the Fe {\sc xv} 284.16 line ($\sim$2.0~MK) because of the high signal-to-noise ratio in these two lines. When measuring the Doppler velocity, we use the average line center over the EIS observing region as the reference wavelength.

We co-align images from different instruments. The AIA image is co-aligned with the HMI magnetogram using the aia\_prep subroutine. The co-alignment between AIA and EIS observations is made using the 193 \AA~ image observed by AIA and 195 \AA\ image observed by EIS.

\section{Heating and Evolution of Two Flare Loops}
\label{result}

High resolution AIA imaging observations allow us to identify flare loops and their footpoints, and study their evolution from the lower-atmosphere to corona. The morphology evolution suggests that the flare is composed of two loop systems that are formed and heated sequentially. In this study, we treat these as two big loops, and identify four footpoints where these two loops are rooted. These four footpoints, named hereafter as FP1, FP2, FP3, and FP4, are indicated in the middle-left panel in Figure 1. They each have an area of $5\arcsec\times5\arcsec$ (limited by the EIS spatial resolution of $5\arcsec\times1\arcsec$). FP1 and FP2 are located in the positive magnetic field region, and FP3 and FP4 are in the negative field region. Clearly shown in the EUV images (the middle-right panel in Figure 1), FP2 and FP3 are connected by a relatively short loop (called LP23 hereafter), and FP1 and FP4 by a longer loop (LP14). In the bottom panels of Figure \ref{obsfig} we show the EIS intensity and Doppler velocity maps in the Fe {\sc xv} line. The four footpoints (and the two flaring loops) are also indicated in these maps.

To study heating and evolution of these two loops, we adopt our recently developed method to infer heating rates from the UV 1600 \AA\ emission at the four footpoints, and then use a 0D loop heating model to compute plasma evolution in the two loops.

\subsection{Evolution of Flare Plasmas with Observationally Inferred Heating Functions}
\label{mod}

A 0D EBTEL model \citep{klim08} is used to compute the coronal response to a specified energy input. The EBTEL model solves time-dependent energy and momentum equations to compute the mean temperature $T$ and density $n$ of coronal plasmas: 
\begin{equation}
 \frac{dn}{dt}=-\frac{2\,c_2}{5\,k_B\,T}\,\left\lbrack\frac{F_c}{L}+c_1\,n^2\,\Lambda(T)\right\rbrack
 \label{eq:n}
\end{equation} 
\begin{equation}
 \frac{dp}{dt}=\frac{2}{3}\,\left\lbrack Q(t)-(1+c_1)\,n^2\,\Lambda(T)\right\rbrack
\label{eq:p}
\end{equation}
where $p=2\,k_B\,n\,T$ is the pressure, $k_B$ is Boltzmann's constant, $c_2$=0.87 is the typical ratio of mean temperature to apex temperature, $L$ is the loop half-length (note that the model assumes a symmetric loop, so only the half-length is considered), $Q(t)$ is the volumetric heating rate, $\Lambda(T)$ is the radiative loss function, and $F_c$ is the thermal conductive flux. Thermal conduction is computed with a classical Spitzer-Harm form, $F_c=-\frac{2}{7}\,\kappa_0\,(T/c_2)^{7/2}/L$, where $\kappa_0=1.0\times10^{-6}$ in cgs units. The constant $c_1$ specifies the radiative loss through the transition region at the footpoint of the half-loop as empirically proportional to the coronal radiative loss. All the constants used in the model, except $c_1$, in this study are taken the same values as in \cite{klim08}.

The model assumes that the coronal plasma in a flare loop is heated directly by an ad-hoc volumetric heating rate $Q(t)$, or indirectly by non-thermal beams that deposit most energy in the lower-atmosphere and then subsequently heat the corona by an upflow (the chromospheric evaporation). The M1.0 flare studied here exhibits a small amount of impulsive X-ray emission above 25 keV energy band for only one minute, so we neglect the non-thermal energy input to the two loops in this study. The possible effect by non-thermal heating will be discussed in Section \ref{discon}. The cooling terms include thermal conduction and radiation. The model also includes an enthalpy flux, or mass flow. The upward enthalpy flux (chromosphere evaporation) plays the role of heating the corona, whereas a downward enthalpy flux (coronal condensation) cools down the corona as an energy loss term. The enthalpy flux transfers energy and mass between the corona and its base at the transition region \citep{brad10a}.

\begin{figure*}
\begin{center}
\includegraphics[height=11cm]{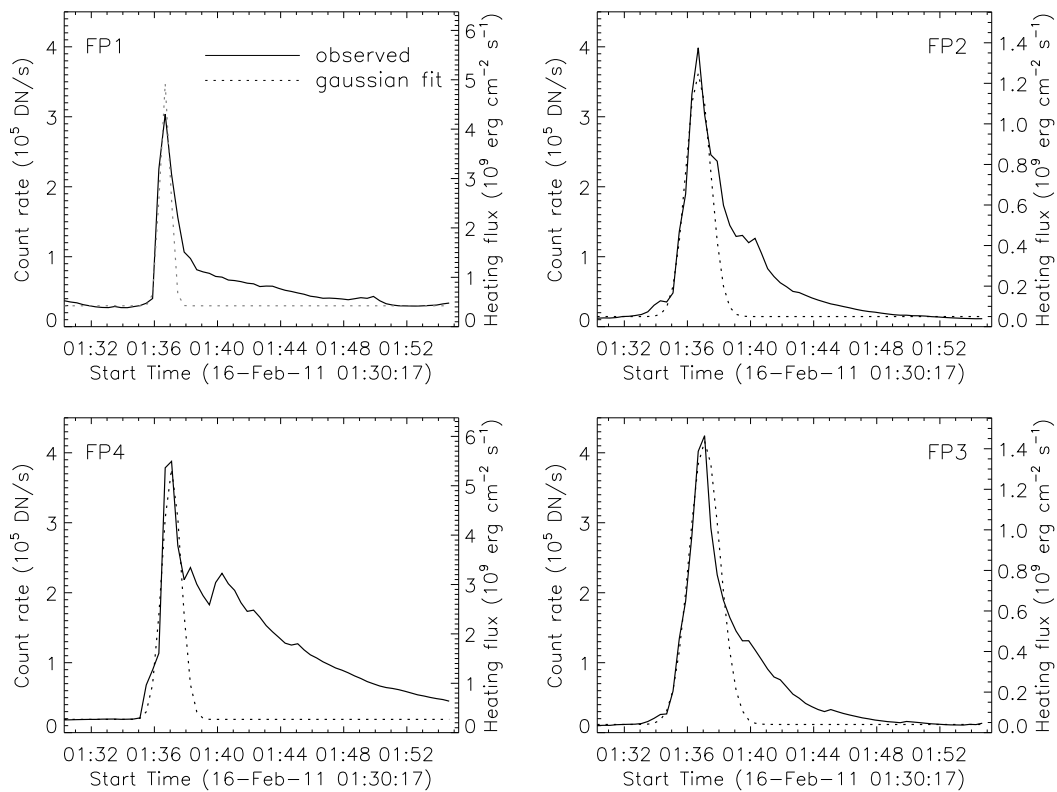}
\caption{Observed UV 1600 \AA~light curves (solid) of the four footpoints. The impulsive rise phase of each light curve is fitted with a half-Gaussian function. Then the symmetric full-Gaussian (dotted) is used to construct the heating function (see text), which corresponds to the right coordinate.}
\label{htffig}
\end{center}
\end{figure*}

Among these terms, the heating rate $Q(t)$ and the loop half-length $L$ are required inputs, and the other terms are calculated from the plasma evolution. For $L$, we assume a semi-circular loop, and its half-length is given by $L=\pi D/4$, where $D$ is the distance between the two footpoints connected by the loop. The heating rate $Q(t)$ is inferred using the method by \cite{qiuj12}, which assumes that the heating flux $H(t) \equiv Q(t) L$ is proportional to the UV emission in the rise phase at the footpoint of the flare loop:
\begin{equation}
H(t)=\lambda I_0\exp\left(\frac{-(t-t_0)^2}{2\,\tau^2}\right)~erg~cm^{-2}~s^{-1}, (0<t<\infty)
\end{equation}
In the equation, $I_0$ is the peak count rate (in units of DN s$^{-1}$) measured by AIA, $t_0$ is the peak time, and $\tau$ is the characteristic rise time. They are determined by fitting the rise phase of the footpoint UV light curve to a half-Gaussian (see Figure \ref{htffig}). The heating rate is also assumed to be symmetric in time, so it is a full Gaussian even though the fitting uses only the rising half of the light curve. The scaling parameter $\lambda$ (in units of erg cm$^{-2}$ DN$^{-1}$) relates the magnitude of the heating flux to the UV peak count rate, and is a free parameter that is adjusted to best match the computed plasma coronal radiation to the observed values.

In this study, we derive $H(t)$ from the four selected footpoints, and therefore model four half-loops (LP1, LP2, LP3, and LP4) rooted at the four footpoints. Since these four half-loops indeed make two full loops, the parameters $L$, $\lambda$, and $c_1$ are the same for the two half-loops LP1 and LP4 rooted at FP1 and FP4, and the same for the two half-loops LP2 and LP3 rooted at FP2 and FP3. The optimal values of $\lambda$ and $c_1$ are chosen from comparison with coronal observations. In addition to the impulsive heating rate $Q(t)$, we also prescribe a background heating term $Q_{bk}$, typically taken as $Q_{bk}=10^{-4}$ erg cm$^{-3}$ s$^{-1}$ \citep{qiuj12}, which is a small fraction of $Q(t)$. The EBTEL computes the plasma evolution with the initial temperature $T_0$ and density $n_0$. Numerical experiments have shown that selection of the initial temperature and density does not modify the plasma evolution once the impulsive heating sets in. We use different initial densities in the loops so that the mean plasma velocity is zero prior to the impulsive heating, namely, the coronal plasma is in static equilibrium before the flare. All these parameters are presented in Table~\ref{partab}.

\begin{table}
\centering
\small
\caption{\textsc{\small{Parameters Used in This Study}}}
\label{partab}
\begin{tabular}{ccccccc}
\tableline
\tableline
\multicolumn{1}{c}{half-loops} & $L$(Mm) & $\lambda$(10$^{4}$ erg~cm$^{-2}$~DN$^{-1}$) & $c_1$ & $Q_{max}$(erg~cm$^{-3}$~s$^{-1}$) & $n_0$(10$^9$ cm$^{-3}$) & $T_0$(MK) \\
\tableline
  LP1 \& LP4 & 28 & 1.4 & 1.2 & 1.6 & 0.5 & 1.0  \\
  LP2 \& LP3 & 12 & 0.3 & 3.5 & 1.2 & 0.7 & 1.0  \\
\tableline
\end{tabular}
\end{table}

\begin{figure*}
\begin{center}
\includegraphics[height=16cm]{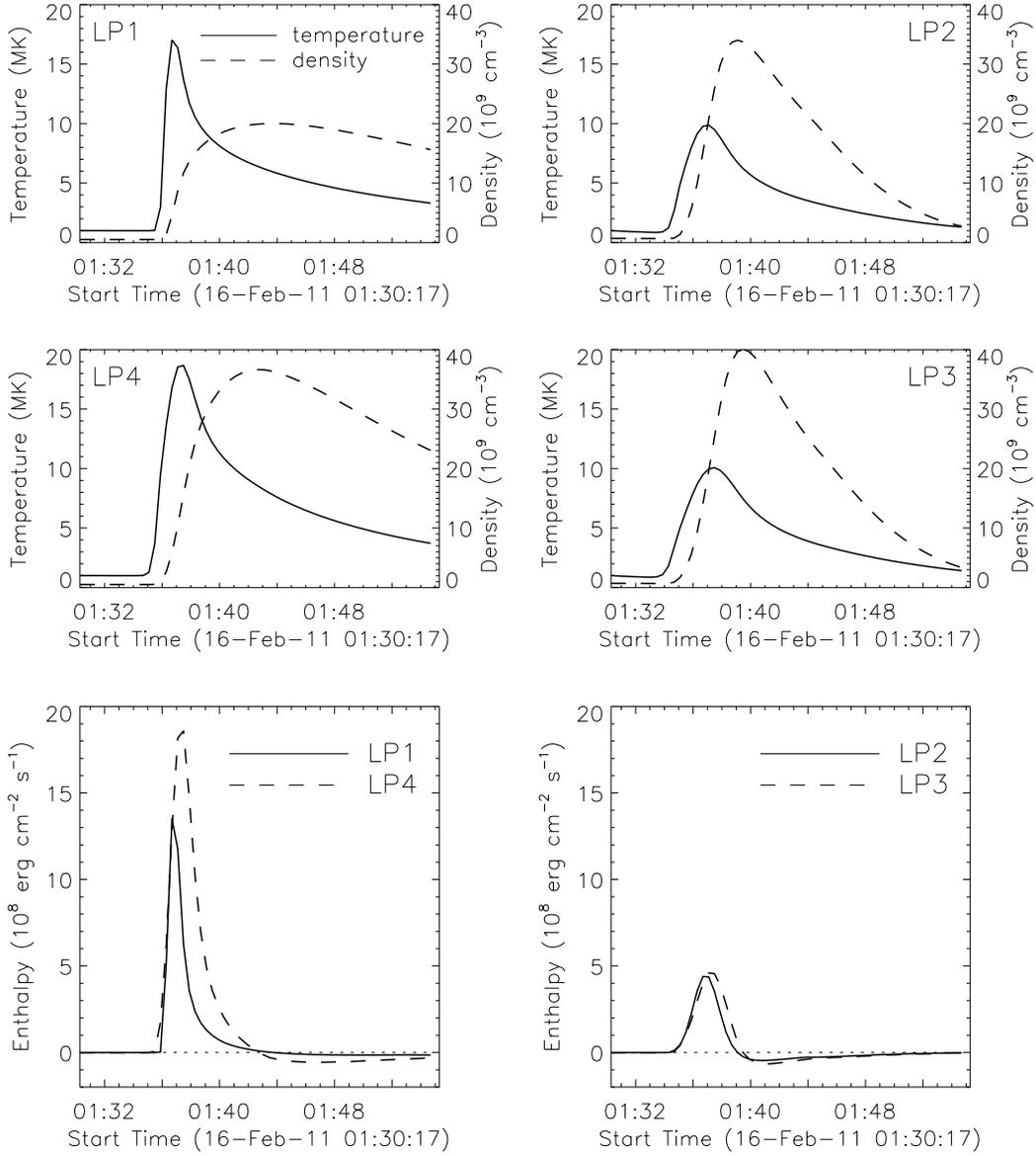}
\caption{The top four panels show computed evolution of the plasma temperature (solid lines) and density (dashed lines) in half-loops rooted at the four footpoints. The bottom two panels plot the mean enthalpy flow in these half-loops.} 
\label{evofig}
\end{center}
\end{figure*}

Using the inputs described above we compute the coronal response by solving the EBTEL equations, (\ref{eq:n}) and (\ref{eq:p}), for each half-loop rooted at a UV-brightened footpoint. Figure \ref{evofig} shows the model results. The top four panels show the evolution of the temperature and density for the four half-loops, and the bottom two panels show the enthalpy flux from the corona through its base, with positive flux indicating upflow and negative flux indicating downflow. Shown in the figure, the coronal temperature rises with the onset of the impulsive heating, and peaks when the heating rate is maximum. Thermal conduction front reaching the coronal base then drives the upflow, or chromospheric evaporation, which raises the density of the coronal loop. The coronal plasma then cools down as dominated by radiative cooling, and in the later cooling phase, the density starts to decrease by coronal condensation, or downflow. 

We note that, for each loop system, although we model two half-loops with heating functions determined from conjugate footpoints separately, the evolution of the two half-loops follow along each other very closely. This is due to the fact that the UV light curves at the two conjugate footpoints are indeed very similar, with comparable rise time (about 2 minutes) and peak count rate. Therefore, the inferred heating rates from conjugate footpoints are nearly identical, resulting in very similar evolution of the two half-loops. Such observations justify using the EBTEL model for this event. 

On the other hand, the short loop and the long loop exhibit different evolution patterns. The short loop is heated for a slightly longer time as derived from the UV rise time, but with a smaller heating rate as determined from the best-fit, in combination with a smaller length scale that allows a greater thermal conduction. As a result, the peak temperature of the short loop is lower than the long loop. The hydrodynamic evolution proceeds faster in the short loop, and the turn-over from upflow to downflow takes place earlier in the short loop at about five minutes after the onset of the impulsive heating. In the long loop, the turn-over takes place at about eight minutes after the heating onset even though the heating time is shorter.

\subsection{Comparison with Observations}
\label{com}

Convolving the computed temperature and density with the instrument response functions, we obtain synthetic coronal radiation fluxes from flare loops to be compared with those observed by AIA and EIS.  For each full loop, we measure EUV fluxes from a series of $5\arcsec\times5\arcsec$ boxes along the length of the loop (see Figure \ref{obsfig}). The measured EUV fluxes from these different locations exhibit nearly identical time profiles, suggesting that the 0D treatment by EBTEL is a close approximation in this case. We sum up all the EUV fluxes from these boxes as the total emission from the loop, and compare it with the computed EUV flux along the full length of the loop. In addition to radiation flux, the model also computes the mean enthalpy flux, which is the difference between conduction flux and radiation flux in the corona. From the enthalpy flux, we can estimate the plasma flow velocity to compare with Doppler velocities measured by EIS at the footpoints of the flare loops. Note that our model only has two free parameters, $\lambda$ and $c_1$, which can be determined by best matching model results with observed emissions in two AIA channels. In this paper, we choose AIA 94 \AA\ and 335 \AA\ for this purpose. Then with this optimal set of parameters, we compare model results with observations from other AIA channels and from EIS to verify validity of the model.

\subsubsection{Comparison with AIA Observations}

\begin{figure*}
\begin{center}
\includegraphics[height=14cm]{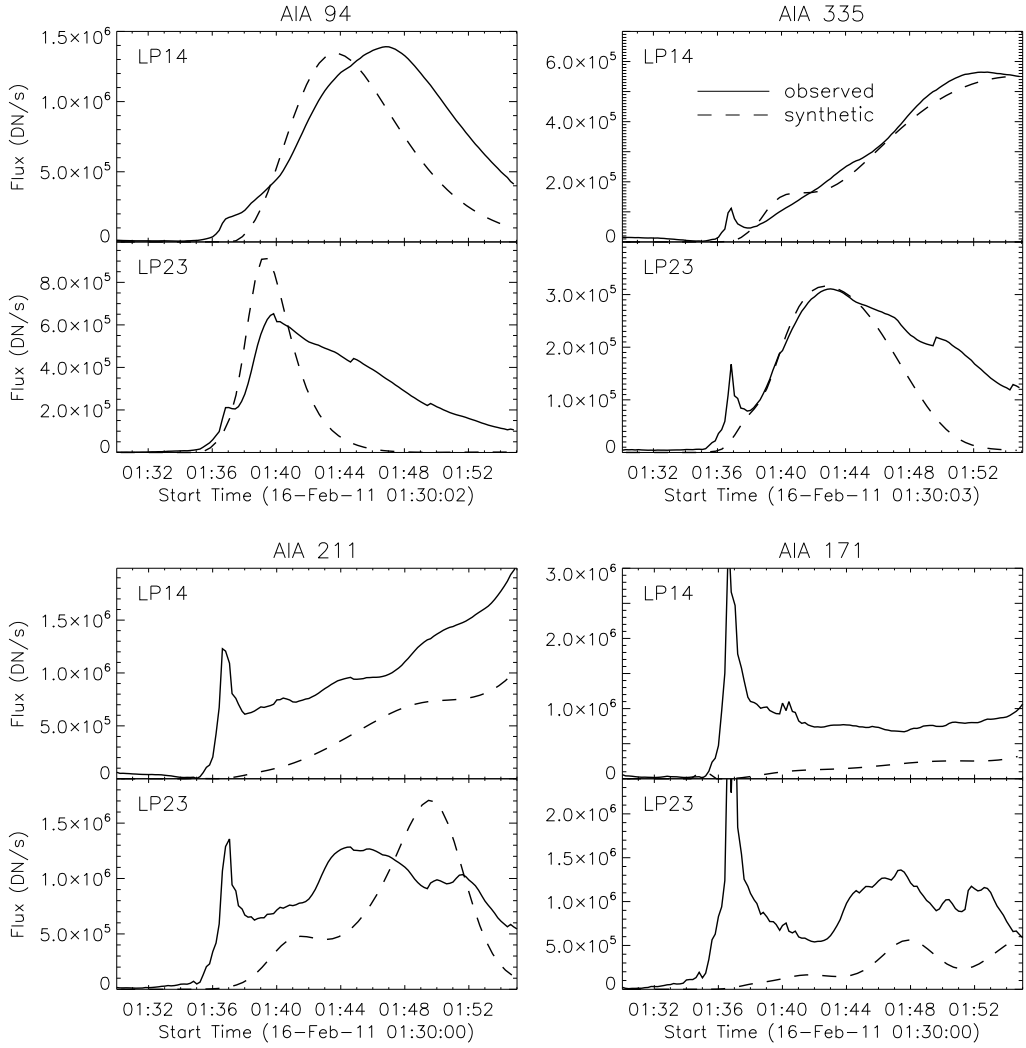}
\caption{Comparison between synthetic (dashed lines) and observed (solid lines) EUV fluxes of the two full loops in 335 \AA, 94 \AA, 211 \AA, and 171 \AA, respectively.}
\label{comAIAfig}
\end{center}
\end{figure*}

Figure \ref{comAIAfig} shows the comparison between computed and observed EUV fluxes in AIA 94 \AA, 335 \AA, 211 \AA, and 171 \AA. The comparison in 193 \AA~and 131 \AA~bands is not shown due to saturation in observed counts in these bands.

The figure shows that, for both loops, the computed EUV fluxes reasonably agree with the observed fluxes in most AIA channels: they all exhibit similar evolution time scales, rise and peak around the same time, and their magnitudes are comparable within a factor of two. Comparing the evolution of the two loops, it is seen that the short loop evolves more quickly, reaching the maximum in higher temperature emission (such as in 94 \AA) earlier than the long loop. The modeled results are basically consistent with the AIA observations.

There are a couple of discrepancies between model and observations, which are likely caused by the limitation of the single temperature approach of the 0D model. Early in the flare, an impulsive emission peak shows up in nearly all EUV bands. This is produced by low-temperature ($<$ 1 MK) plasmas at the footpoint upon impulsive heating at the lower atmosphere \citep{bros12}. In the late phase of the flare loop evolution, the observed abundant EUV flux in a few bands (such as in 94 \AA) is likely produced by cooled plasma at lower temperatures of $<1$ MK. Since the 0D EBTEL model does not reproduce temperature distribution along the loop at a given time, it cannot capture emissions at temperatures beyond the averaged value, which are observed by AIA due to its broad temperature response. This may also cause deviation of peak time (e.g., in 94 \AA~for the long loop). Despite these discrepancies, the overall agreement between model and observations suggests that our method reasonably describes mean properties of heating and evolution of flare loops. 

\subsubsection{Comparison with EIS Observations}

\begin{figure*}
\begin{center}
\includegraphics[height=8cm]{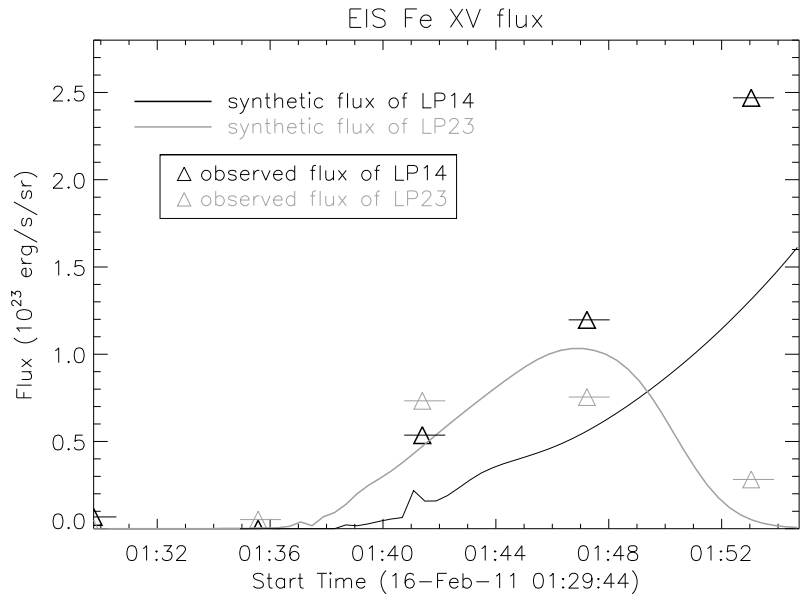}
\caption{Comparison between synthetic (lines) and observed (triangles) fluxes of the two full loops in the Fe {\sc xv} line. The horizontal bars at triangles represent the time ranges of each EIS scan.}
\label{comEISIfig}
\end{center}
\end{figure*}

Comparison is also made with EIS observations. Figure \ref{comEISIfig} shows the synthetic and observed emission fluxes in the EIS Fe {\sc xv} line, which typically forms at 2 MK. It is evident that, in both loops, the computed flux tracks the observed flux, with the short loop evolving more quickly than the long loop. The absolute magnitude of the observed and computed fluxes are comparable by within a factor of two. 

The result for the Fe {\sc xii} line is not shown. The computed mean temperature is higher than the formation temperature of this line, hence it does not produce significant emission in this line. In reality, there is a temperature distribution (namely, the differential emission measure) in the flare loop, the low temperature portion of which would produce emission in this line. Such cannot be described by the 0D EBTEL model.

\begin{figure*}
\begin{center}
\includegraphics[height=13cm]{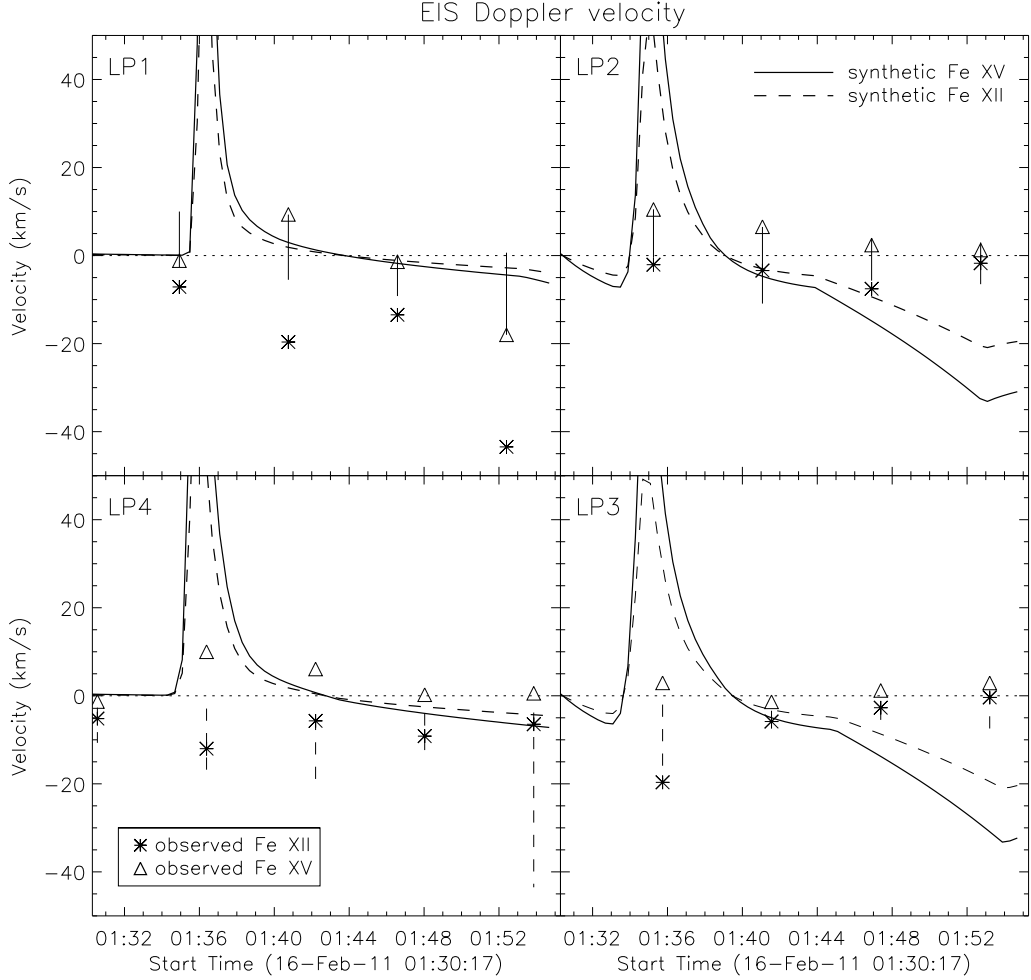}
\caption{Comparison between computed equivalent velocities from the enthalpy flows (see text; solid and dashed lines) and measured Doppler velocities (triangle and asterisk  symbols) at the four footpoints in the Fe {\sc xv} and Fe {\sc xii} spectral lines. The vertical lines at symbols represent the observed velocity ranges along the full loop. For clarity, we plot the velocity range in the Fe {\sc xv} line only at FPs 1 and 2, while in the Fe {\sc xii} line only at FPs 3 and 4.}
\label{comEISVfig}
\end{center}
\end{figure*}

The spectroscopic capability of EIS allows us to further compare the modeled mean flow velocity with the Doppler shift measurements to gain insight in the hydrodynamic behavior of coronal plasma. Figure \ref{comEISVfig} shows the Doppler velocities in the Fe {\sc xii} and Fe {\sc xv} lines measured at the loop footpoints, with the vertical bar marking the range of velocities measured along the full loop. The measurements show that, in the impulsive phase, the hot Fe {\sc xv} line is blue-shifted (positive velocity) and the cooler Fe {\sc xii} line is red-shifted (negative velocity), which may be explained by the momentum balance between upflow (evaporation) and downflow (condensation) upon impulsive heating of the flare atmosphere \citep{canf90}. During the decay, both lines are red-shifted, suggesting that coronal condensation dominates during the cooling phase.

EBTEL computes the mean enthalpy flux $\zeta \propto Pv$, which is the difference between the conductive flux and radiative flux in the corona. This mean-property approach cannot reproduce the velocity profiles along the loop and at different temperatures during the impulsive heating phase, but can be used to estimate the mean flow velocity in the corona. We assume that the coronal loop has a steady-flow, $n(T)v(T) = n_0v_0$, and is in dynamic equilibrium $P(T) = 2\,k_B\,n(T)\,T = P_0$, where $n_0$, $P_0$, and $v_0$ are EBTEL computed mean density, pressure, and velocity of the coronal plasma, respectively. This yields the estimate of the flow velocity $v(T)$ at a given temperature. 

Figure \ref{comEISVfig} shows the flow velocities estimated at the formation temperatures of the Fe {\sc xii} and Fe {\sc xv} lines in comparison with the measurements. It is seen that the model and observations qualitatively agree with each other in the higher temperature Fe {\sc xv} line, both showing the upflow during the impulsive phase and downflow during the decay phase. Specifically, both model and observations show that the turn-over from upflow to downflow occurs around the same time. It occurs after 1:40 UT for the long loop and before 1:40 UT for the short loop.

We note that the magnitudes of the modeled and measured velocities are not directly comparable especially during the impulsive phase, when the steady-flow and equilibrium assumptions are questionable, and the plasma properties along the loop may largely deviate from the mean values computed by EBTEL. Furthermore, the EIS scans missed most part of the impulsive phase, when upflow of high temperature plasmas may be predominant. Therefore, during the impulsive phase, the comparison is rather poor between the EBTEL modeled flow and the observed flow in the two relatively low-temperature lines.

\section{Summary and Discussions}
\label{discon}

We analyze and model two distinctive loop systems that are treated as two big loops in an M1.0 flare. Using the empirical heating functions inferred from observed UV light curves, we compute the evolution of plasmas in these loops via EBTEL. With the model result we reproduce the EUV synthetic fluxes and compare them with observations by AIA and EIS for each of the two loops. We also compare the computed mean enthalpy flow velocity with the measured Doppler velocity from EIS. The model results agree with observations in many aspects. The two different loops show different coronal plasma evolution patterns, with the short loop evolving more quickly than the long loop. These patterns are revealed in observations and as well reproduced in the model. Our results suggest that the observationally constrained heating rates and the EBTEL modeling with only two free parameters provide a reasonable description of the mean properties of heating and evolution of flare loops.

In our model, the loop length, and the timing and duration of the heating rate are derived directly from observations. Previous study using the similar method uses exactly the same set of parameters to scale the heating rate and energy loss rate for a few thousand loops. Our experiment of modeling two spatially resolved loop systems then allows us to investigate the variations of these model parameters  $\lambda$ and $c_1$ in different loops. The parameter $\lambda$ relates the UV emission counts directly from data to the amount of direct heating in the corona ($QL$), and $c_1$ scales the total energy loss through the transition region, or the base of the flare loop. Our results show that, to best match the mean properties of loops, different parameters have to be used. The best-match $\lambda$ for the long loop is four times that of the short loop (see Table \ref{partab}). On the other hand, $c_1$ for the long loop is one third that of the short loop. For the long loop, the value of $c_1$ is comparable with the average value determined in \cite{qiuj12} who studied a few thousand long loops; and for the short loop, the optimal $c_1$ value is substantially larger. These differences may indicate different heating mechanisms and their subsequent low-atmosphere responses in the two loops. A plausible explanation for these differences may involve heating by electron beams. Imaging observations by {\em RHESSI} suggest that during the early impulsive phase, there are hard X-ray sources located at the feet of the short loop. {\em In-situ} heating by non-thermal electrons produces significant UV emission at these locations (\citealt{allr05, cheng10, cheng12}, and references therein). Therefore, less amount of UV emission is contributed by direct heating in the corona, which heats the lower atmosphere through thermal conduction during the impulsive phase. A semi-quantitative analysis including non-thermal heating term in the EBTEL modeling is given by \cite{liuw12}. A larger $c_1$ value used for the short loop also indicates a greater amount of radiative loss through the transition region than in long loops. This suggests that $c_1$ might not be an independent constant but may vary with the loop dynamics \citep{carg12}.

Flow velocity is an important property in studying the dynamic evolution of flares. Spectroscopic velocity measurements can help diagnose the process of energy release and plasma evolution. During the impulsive phase, chromospheric evaporation (upflow), driven either by electron beams or thermal conduction \citep{acto82, fish85a, fish85b, bros04, mill08}, provides indirect heating apart from the direct heating ($Q$) in the corona. In the cooling phase, coronal draining (downflow) dominates. In our present study, the spectroscopic observations do not adequately cover the impulsive phase of the flare, but our analysis shows that the modeled timing of the turn-over from upflow to downflow is consistent with the measured Doppler velocity in the decay phase, which supports the result of \cite{brad10a,brad10b}. Our analysis shows the potential of using Doppler shift measurements to constrain the EBTEL model that invokes enthalpy flow in the flare evolution. In our future investigation, the high efficiency of the EBTEL model will allow us to model a good number of spatially resolved loops, and statistically investigate the relationship between energy release and plasma flow. 

Despite its advantages, the EBTEL model does not reproduce the temperature distribution along the loop, which may explain the insufficient low temperature plasma emission from the model, as compared with observations. Furthermore, recent studies suggest that the observed evaporation velocity during the impulsive phase is temperature dependent \citep{mill09, chen10, ying11}, which requires more sophisticated models, such as 1D hydrodynamic model, to compute the plasma evolution especially during the impulsive heating phase. The temporal properties of heating rates determined in this study can be used as the reference for 1D models; on the other hand, a 1D model includes more assumptions or free parameters such as those describing distribution of heating along the loop and therefore invokes addititional observational constraints.

\acknowledgments
Y.L. would like to thank Dana Longcope and Wenjuan Liu for help with EBTEL modeling and scientific discussion, and the anonymous referee for valuable comments. Y.L. and M.D.D are supported by NSFC under grants 10878002, 10933003, and NKBRSF under grant 2011CB811402. Y.L. is also supported by NSFC under grant 11103008 and  CSC under file No. 2011619032. J.Q. is supported by NSF grant ATM-0748428. \textit{SDO} is a mission of NASA's Living With a Star Program. \textit{Hinode} is a Japanese mission developed and launched by ISAS/JAXA, collaborating with NAOJ as a domestic partner, and NASA (USA) and STFC (UK) as international partners. Scientific operation of the \textit{Hinode} mission is conducted by the \textit{Hinode} science team organized at ISAS/JAXA. Support for the post-launch operation is provided by JAXA and NAOJ (Japan), STFC (UK), NASA, ESA, and NSC (Norway).

\bibliographystyle{apj}


\end{document}